\documentclass[%
 reprint,
superscriptaddress,
showpacs,
showkeys,
 amsmath,amssymb,
 aps,
 prl,
]{revtex4-1}

\usepackage{graphicx}
\usepackage{dcolumn}
\usepackage{bm}
\usepackage{hyperref}

\begin{document}


\title{Absence of proton tunneling during the hydrogen bond symmetrization in $\delta$-AlOOH}
\author{Florian Trybel}
\email{florian.trybel@liu.se}
\affiliation{ Bayerisches Geoinstitut, Universit{\"a}t Bayreuth, D-95440 Bayreuth, Germany}
\affiliation{ Department of Physics, Chemistry and Biology (IFM), Link{\"o}ping University, SE-581 83, Link{\"o}ping, Sweden} 
 \author{Thomas Meier}
\affiliation{ Bayerisches Geoinstitut, Universit{\"a}t Bayreuth, D-95440 Bayreuth, Germany}
\affiliation{Center for High Pressure Science and Technology Advanced Research (HPSTAR), Beijing 100094, China}
\author{Biao Wang}
\affiliation{ Bayerisches Geoinstitut, Universit{\"a}t Bayreuth, D-95440 Bayreuth, Germany}
\affiliation{ Department of Earth Sciences, University of Oxford, OX1 3AN Oxford, United Kingdom}
\author{Gerd Steinle-Neumann}%
\affiliation{ Bayerisches Geoinstitut, Universit{\"a}t Bayreuth, D-95440 Bayreuth, Germany}

\date{\today}

\begin{abstract} 
 $\delta$-AlOOH is of significant crystallochemical interest due to a subtle structural transition near 10 GPa from a $P2_1nm$ to a $Pnnm$ structure, the nature and origin of hydrogen disorder, the symmetrization of the O-H$\cdots$O hydrogen bond and their interplay. We perform a series of density functional theory based simulations  in combination with high-pressure nuclear magnetic resonance experiments on $\delta$-AlOOH up to 40 GPa with the goal to better characterize the hydrogen potential and therefore the nature of hydrogen disorder. Simulations predict a phase transition in agreement with our nuclear magnetic resonance experiments at $10-11$ GPa and hydrogen bond symmetrization at $14.7$ GPa. Calculated hydrogen potentials do not show any double-well character and there is no evidence for proton tunneling in our nuclear magnetic resonance data.
\end{abstract}

\maketitle

\section{Introduction}
Hydrogen is an important chemical component in the Earth's mantle, as even a small amount can strongly affect key properties of minerals, such as melting temperature, rheology, electrical conductivity and atomic diffusion \cite{bercovici2003,inoue1994,karato1986,yoshino2006}. Therefore, over the past 20 years, many hydrous minerals, such as dense hydrous magnesium silicates \cite{thompson1992}, have been synthesized at high-pressure ($P$) and high-temperature ($T$) conditions and investigated as potential candidates for hydrogen transport to the lower mantle.
However, most of these minerals decompose at $P < 60$ GPa, where phase H breaks down to MgSiO$_3$ bridgmanite and a fluid component \cite{ohtani2014stability,tsuchiya2019first,nishi2018thermal}.

In 2017, AlSiO$_3$(OH) was found in diamond inclusions \cite{wirth2007} from the mantle transition zone at a depth of 410-660 km. High-$P$, high-$T$ experiments revealed that this phase can form from hydrous sediment components at upper mantle conditions (10 - 12 GPa) \cite{ono1998}, and decomposes to $\delta$-AlOOH and SiO$_2$ stishovite at conditions similar to those found at the base of the mantle transition zone ($P>$20 GPa) \cite{sano2004}. $\delta$-AlOOH is particularly interesting as it shows a wide stability range including conditions along the geotherm of a subducting slab \cite{sano2008,duan2018phase,su2021effect} and may therefore be a potential host of hydrogen in Earth's lower mantle.

\begin{figure}[h!t!]
\includegraphics[width=0.50\textwidth]{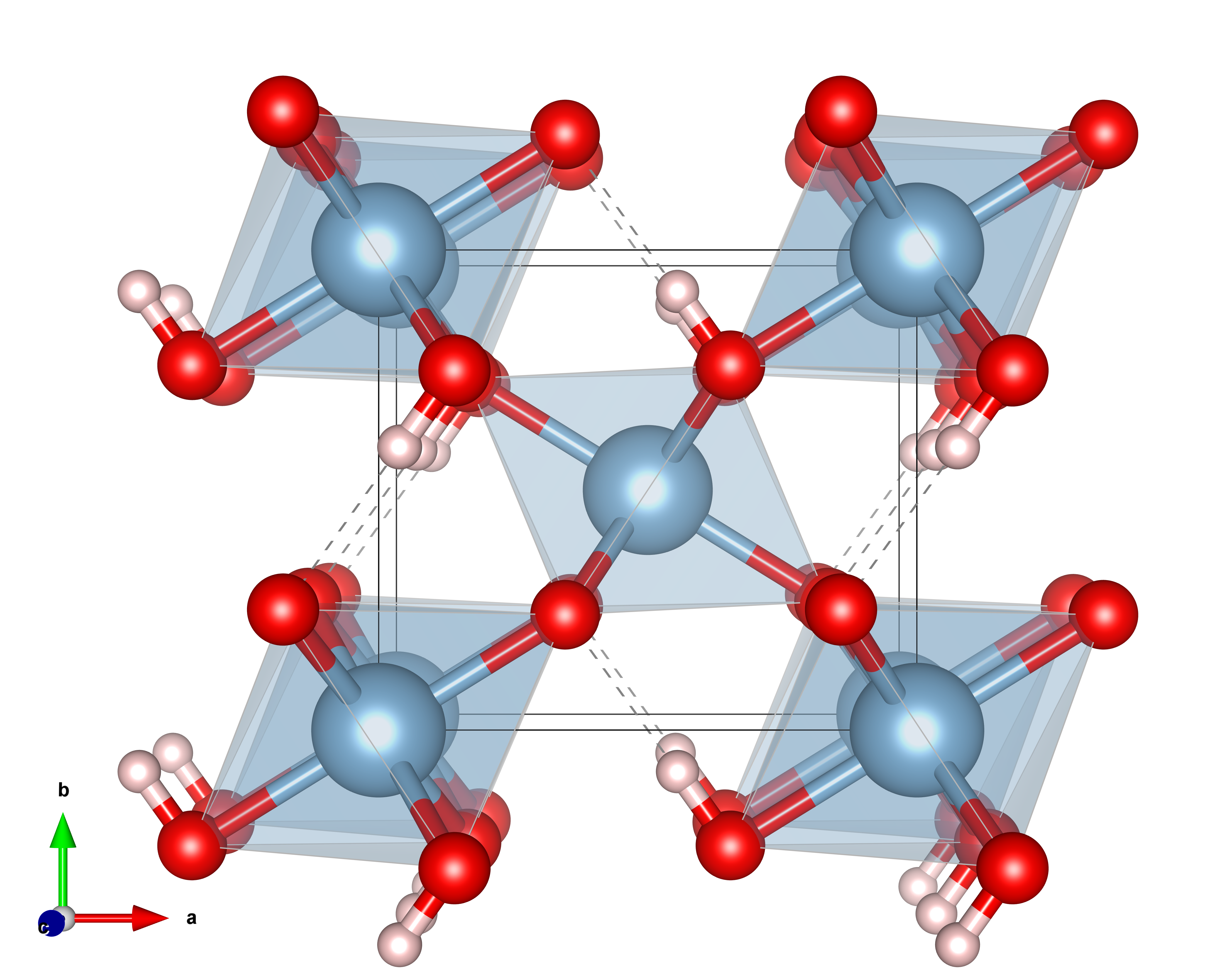} 
\caption{Crystal structure of $P2_1nm$ $\delta$-AlOOH. Aluminum is shown in blue, oxygen in red and hydrogen in white. Offset of the hydrogen positions from the center is exaggerated for illustration.}
\label{fig:P21nm_structure}
\end{figure}

$\delta$-AlOOH crystallizes in a primitive orthorhombic lattice with space group $P2_1nm$ at ambient conditions \cite{K2006,ohtani2001,sano2008,suzuki2000}. Its structure corresponds to distorted rutile, with Al and O atoms located on mirror planes (Fig.~\ref{fig:P21nm_structure}). AlO$_6$ octahedra share edges along the $c$ axis and these octahedra chains are connected via corners (Fig.~\ref{fig:P21nm_structure}). There are two distinct oxygen positions (O1 and O2) at the vertices and in the equatorial plane, respectively. Layers of AlO$_6$ octahedra oriented in different directions are connected with an asymmetric hydrogen bond between the layers. 

Single-crystal synchrotron X-ray diffraction (XRD) \cite{K2006,kuribayashi2014} and neutron diffraction measurements \cite{SF2008,SF2018} identified a sub- to super-group phase transition from $P2_1nm$ to $Pnnm$ at $\sim10$ GPa where O1 and O2 positions become equivalent. The transition is accompanied by a rotation of the AlO$_6$ octahedra  by $\sim1^{\circ}$, and  in the same $P$ range the hydrogen position is predicted to symmetrize \cite{SF2018,kuribayashi2014,pillai2018,cortona2017,tsuchiya2002}. In Fourier difference maps calculated from their neutron diffraction data, \citet{SF2018} found a bimodal hydrogen distribution between 9.5 and 18 GPa, supporting previous suggestions  \cite{kuribayashi2014,cortona2017,pillai2018} that a double-well potential along the diagonal O-O direction may exist, which could give rise to proton tunneling, similar to the state found during the ice-VII to ice-X transition \cite{lin2011correlated,meier2018,trybel2020proton}. No direct evidence of tunneling has been found to date, however.

We investigate the phase transition, hydrogen bond symmetrization (a central uni-modal proton distribution between the two respective oxygen atoms) and the possibility of proton tunneling in $\delta$-AlOOH, combining density functional theory (DFT) based calculations and high and low field high-$P$ nuclear magnetic resonance (NMR) spectroscopy. With DFT, we perform a stepwise optimization of the host lattice and the hydrogen positions over a wide volume ($V$) range and analyze the potential seen by the hydrogen atoms as well as the geometry of the AlO$_6$ octahedra. We analyze the signal shift as well as the full width at half maximum (FWHM) of the high-field NMR experiments at $P$ up  to 40 GPa (\textit{c.f.}  Fig. S1 of the supplementary material \cite{Supplement}), searching for characteristic features of a phase transition, and use low-field NMR data at 5.6 GPa to investigate indications of proton tunneling \cite{meier2018}.
\begin{figure}[h!]
\includegraphics[width=0.48\textwidth]{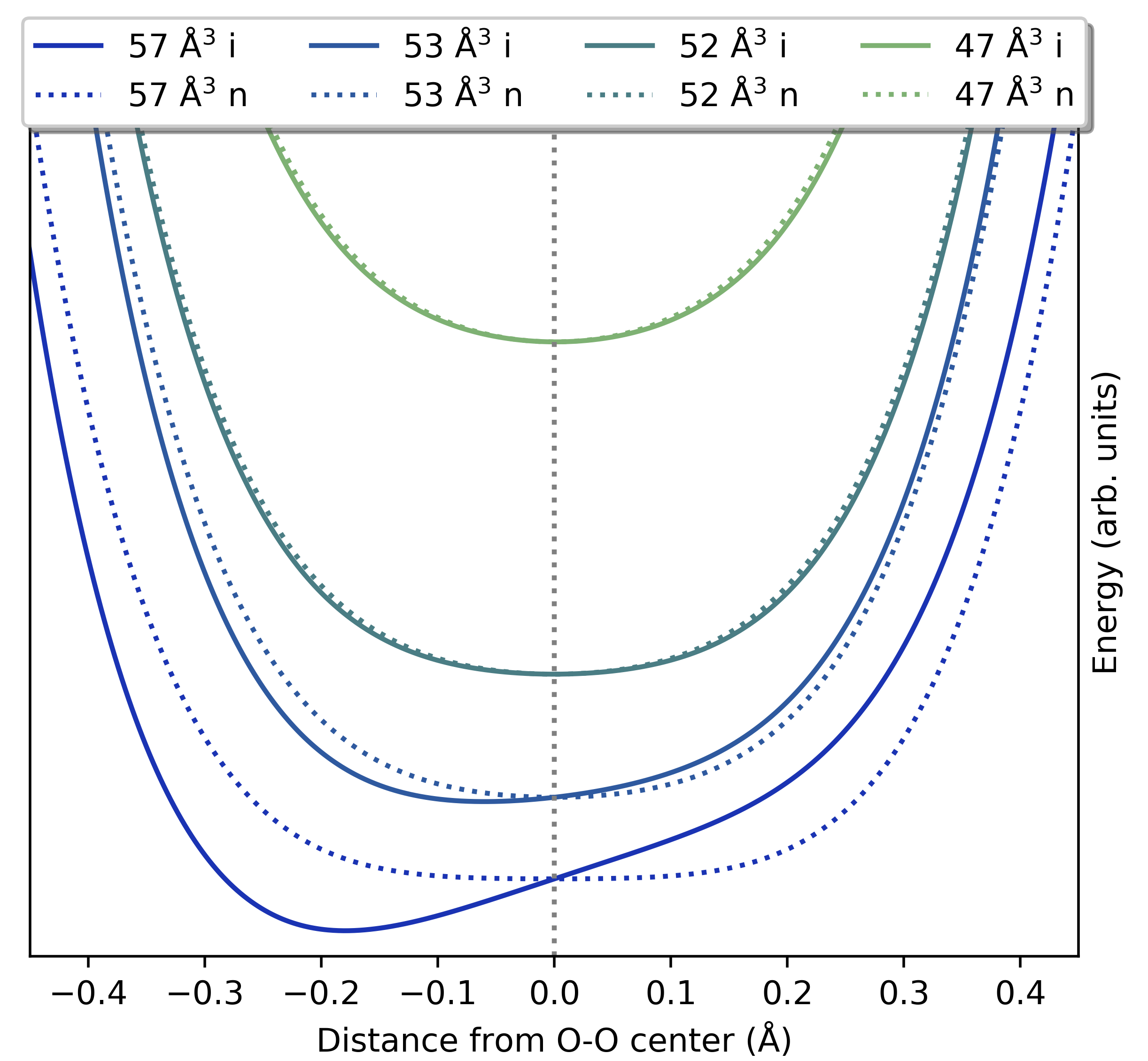} 
\caption{Comparison of normal (dotted) and inverse (solid) sampling (Fig. S2 in the supplementary material \cite{Supplement}) of the potential seen by the hydrogen atom. At large volumes (low $P$), normal sampling leads to a symmetric and inverse sampling to an asymmetric single-well potential with a significantly lower energy ($\sim$ 0.2 eV). With increasing compression, this difference vanishes as the two oxygen positions become equivalent. By construction, the potentials are equivalent at the central position (indicated by the vertical dashed line).}
\label{fig:PotSampling}
\end{figure}

\section{Computational Details}

All DFT simulations are performed using the Quantum Espresso package \cite{QE-2009,QE-2017}, where possible GPU-accelerated \citep{romero2017performance}. We use projector augmented wave atomic files for Al, O and H based on the PBEsol \cite{perdew2008} approximation to exchange-correlation which was previously found to show excellent agreement with experimental data \cite{cortona2017}. For Al the 2p electrons and lower and for O the 1s electrons are treated as semi-relativistic core states. Convergence tests with a threshold of $10^{-5}$ Ry/atom lead to a Monkhorst-Pack k-point grid \cite{MonkhorstPack} of 12$\times$8$\times$8 for primitive unit cells and a cutoff energy for the plane wave expansion of 140 Ry. We use the FINDSYM code \cite{stokes2005} for symmetry analysis and VESTA 3 \cite{Vesta} for visualization of structures.

\section{Experimental Details}
High-$P$ cells with pairs of 250 $\mu$m culet diamond anvils are used, and the preparation of the NMR experiments closely follows the procedure outlined in our previous work \cite{ meier2020proton}.  3.2 mm excitation coils are formed from single turn cover inductors made from 50 $\mu$m copper-coated teflon foil. The diamond anvil is coated with 1 $\mu$m of copper using physical vapor deposition, and subsequently Lenz lens resonators are cut out of this layer using focused ion beam milling. 

A $~20~ \mu$m$^3$ crystal of $\delta$-AlOOH (synthesis described in \cite{simonova2020}) is placed in the sample chamber and the DAC is filled with neon as a $P$-transmitting medium. Both excitation coils are mounted central to the diamond anvils and connected in a Helmholtz coil arrangement after closing the cell. For the high-field measurements, 1045 mT with a corresponding $^1$H resonance frequency of $\sim$ 45 MHz in an electromagnet is used. Additional homonuclear $^1$H-$^1$H decoupling experiments are conducted using a Lee-Goldburg \cite{meier2019} saturation pulse of 25 W prior to the spin excitation to obtain high resolution $^1$H-NMR spectra. An additional DAC prepared in a similar manner is filled with distilled H$_2$O and used as a resonance shift reference. Low-field measurements are performed at 125 mT and a resonance frequency of $\sim$ 5 MHz, using the same electromagnet. Resulting spectra are analyzed by line form matching to the experimental signal \cite{meier2020hydro}. Pressure is calculated from the Raman signal of the diamonds \cite{akahama2004high,akahama2006pressure} and the AlOOH volume using the equation-of-state (EOS) from \citet{simonova2020}. 
\begin{figure}[h!]
\includegraphics[width=0.45\textwidth]{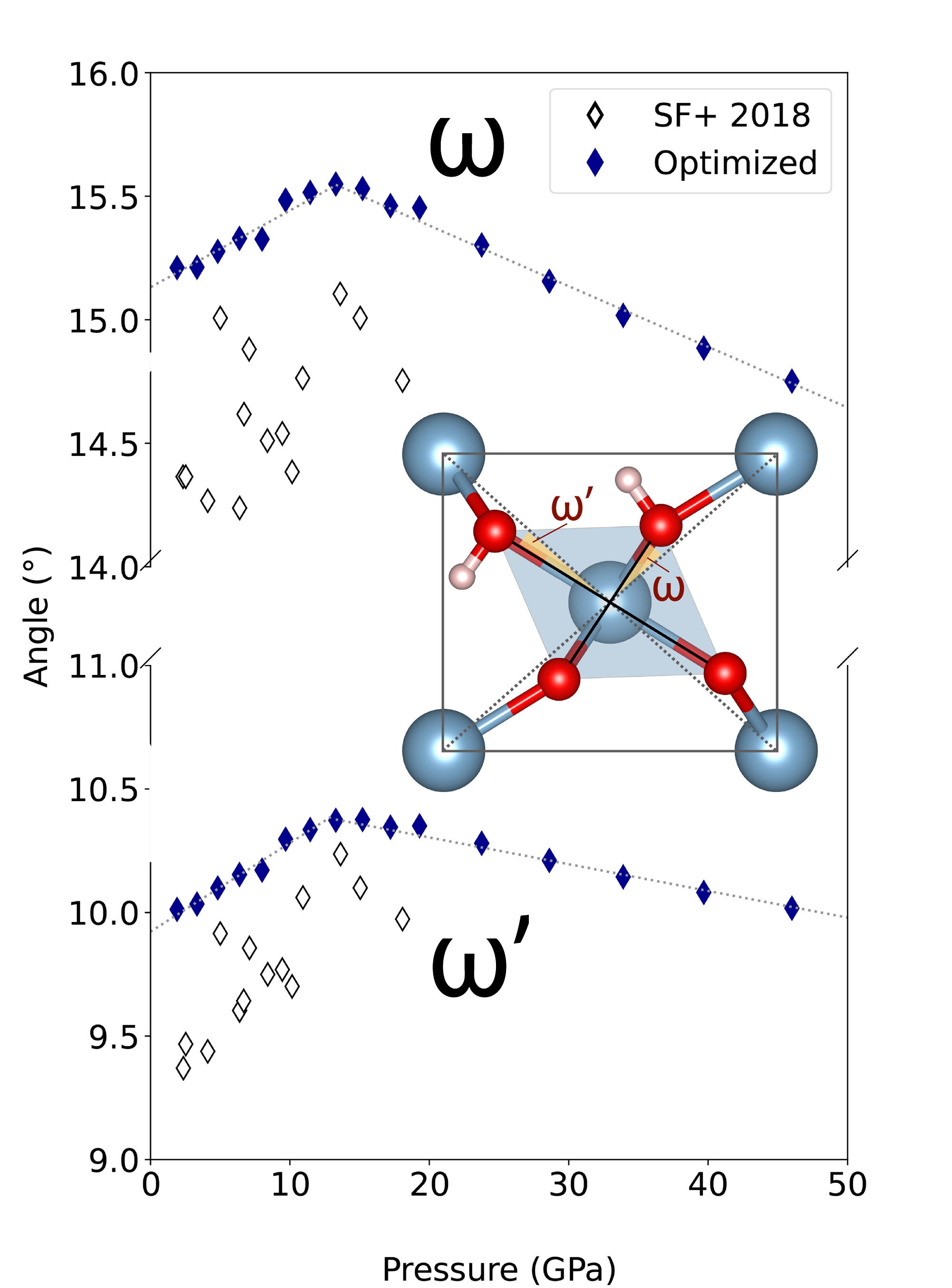} 
\caption{Angle analysis for the fully relaxed cell (filled symbols) and the experimental values by \citet{SF2018}  (open symbols, SF+2018) following their definition (inset).}
\label{fig:Angles}
\end{figure}

\section{Structural Optimization \& Hydrogen Potential}

In the computations, we start structural optimization with reported experimental low-$P$ structures from \citet{K2006} and \citet{SF2018} and optimize the hydrogen position for $V$ between $57$ and $47$ \AA$^3$. We sample and optimize the hydrogen positions along the diagonal oxygen-oxygen direction first, relax the coordinates of all atoms, perform a second sampling and optimization of the hydrogen positions, relax the cell parameters and perform a final sampling and optimization of the hydrogen positions. Both initial structures \cite{K2006,SF2018}  converge to the same coordinates within the first steps of the relaxation. 

We sample configurations by displacing the hydrogen atoms along the diagonal oxygen-oxygen direction in the $a$-$b$ plane by starting with both hydrogen atoms being close to the respective oxygen atom with the smaller $b$ coordinate (normal configuration) and from the configuration shown in Fig.~\ref{fig:P21nm_structure} (inverse configuration), \textit{c.f.}  Fig. S2 of the supplementary material \cite{Supplement}. From the energy obtained in each sampling step, we construct the potential seen by the hydrogen atom, similar to our previous work on the ice-VII to ice-X transition \cite{trybel2020proton} via spline interpolation at each sampled $V$ for both configurations. The final positions of the hydrogen atoms are obtained as the minima of the respective spline interpolation (Fig.~\ref{fig:PotSampling}). 

We find distinctly different potential symmetries for normal and inverse sampling with respect to the center of the diagonal oxygen-oxygen distance: normal sampling results in a symmetric potential, inverse sampling in an asymmetric potential ($\sim0.2$ eV lower in energy).  Under compression the potentials become narrower in both sampling types, and for inverse sampling asymmetry decreases. At $V \lesssim 52.4\pm 0.1$ \AA$^3$ the potential obtained by inverse sampling becomes symmetric and the energy difference between the two potentials approaches zero, with the inverse sampling remaining slightly lower in energy over the full $V$ range.

Contradicting prior suggestions \cite{cortona2017,pillai2018}, both potentials do not show any double-well character, even though calculating the potential seen by the hydrogen atom from Kohn-Sham DFT should strongly overestimate the potential well without further consideration of the quantum nature of the hydrogen atoms \cite{cortona2017,trybel2020proton,drechsel2014quantum}. 

After each optimization step, we analyze the space group of the resulting cell and track the rotation of AlO$_6$ octahedra using angles $\omega$ and $\omega^{'}$ as defined by \citet{SF2018} as a function of compression (Fig.~\ref{fig:Angles}).

We find an increase in the angle of $\sim0.6 ^{\circ}$ when compressing from 57 \AA$^3$ to  $\sim 53$ \AA$^3$, followed by a decrease at higher compression for both $\omega$ and $\omega^{'}$. The angles are in general $\sim 0.5^{\circ}$ larger than the experimental values by \citet{SF2018} which show large scatter; the difference between $\omega$ and $\omega^{'}$ and the $P$ dependence they report are in very good agreement with our calculations.

\section{Nuclear Magnetic Resonance Spectroscopy}

\begin{figure*}[th!]
\includegraphics[width=\textwidth]{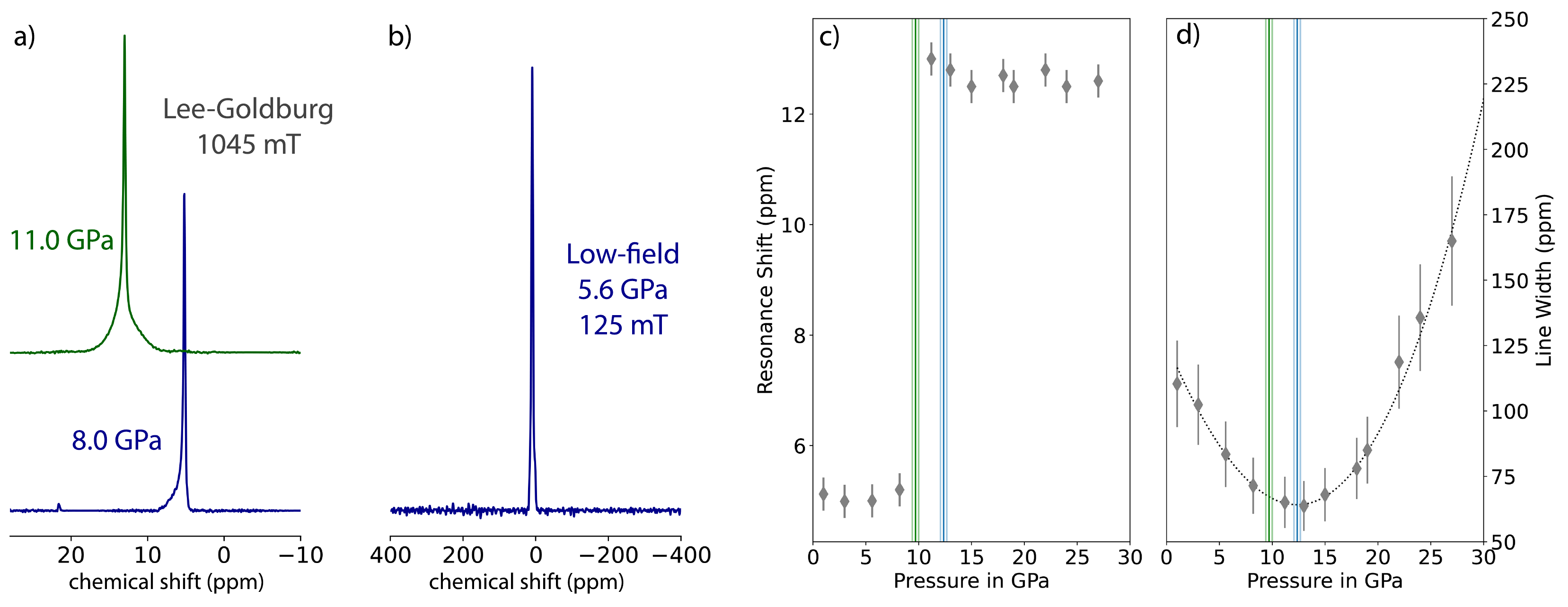} 
\caption{\textbf{(a)} Lee-Goldburg $^1$H-NMR spectra at 8 and 11 GPa at a field of 1045 mT with H$_{\rm 2} $O as a chemical shift reference. \textbf{(b)} NMR spectrum of $\delta$-AlOOH at $5.6$ GPa (shifted to zero) at a field of 125 mT. 
\textbf{(c)} Shift of the resonance as a function of pressure in the Lee-Goldburg spectra. The shift increases abruptly at $P=9.9 \pm 0.6 $ GPa  from $\sim 5$ ppm to $>12$ ppm. \textbf{(d)} Full width at half maximum of the resonance peak as a function of $P$. A minimum in line width occurs at $ P=12.7 \pm 0.4$ GPa ($ \sim 52.7$\AA$^3$). The dotted black  line is a third order polynomial fit to the data between 5 and 20 GPa. In panels (c) and (d), the blue vertical lines shows the minima of the polynomial fit to the line width data and the green line indicates the jump in the resonance shift with the respective errors indicated by the dotted lines.}
\label{fig:NMR_Spectra}
\end{figure*}

NMR experiments employing Lee-Goldburg decoupling pulses lead to line widths of $\sim1.5$ ppm (shown for 8 and 11 GPa in Fig.~\ref{fig:NMR_Spectra}a) which permits the analysis of chemical shifts with $\sim 10$ ppm (Fig.~\ref{fig:NMR_Spectra}c). For the chemical shift, we find an abrupt change at $P=9.9 \pm 0.6 $ GPa  from $\sim 5$ ppm to $>12$ ppm, indicating a structural change in the chemical environment of the hydrogen atoms as expected for a phase transition.  We find a single proton signal over the full pressure range indicating a unique, geometrically well defined proton position, in agreement with an earlier NMR study at ambient conditions by \citet{Xue2007}.

Lee-Goldburg decoupling, while retaining information about the isotropic chemical shift interactions and strongly focusing the line width, leads to a cancellation of the non-secular parts of the total spin Hamiltonian and thus to a loss of information about spin interactions due to chemical shift anisotropy, direct homo- and heteronuclear dipole-dipole as well as first order quadrupolar interactions. Therefore, we additionally analyze the line shape of $^1$H-NMR solid-echoes (Fig.~S1 of the supplementary material \cite{Supplement}) and find a minimum in the line width at $P=12.7 \pm 0.4$ GPa (Fig.~\ref{fig:NMR_Spectra}d), indicating a change in proton mobility.

Proton tunneling should lead to a zero field splitting accompanied by detectable tunneling side bands as it introduces an exchange between allowed magnetic transitions with $\Delta m = 1$ and usually forbidden transitions \cite{clough1985}. In work on the ice-VII to ice-X transition in high-$P$ H$_2$O \citep{meier2018}, we showed that such tunneling side bands can be resolved at high $P$ using low-field NMR, and that NMR is sensitive to changes in the tunneling rate during compression. We therefore employ the low-field setup at $P=5.6$ GPa (Fig.~\ref{fig:NMR_Spectra}b) and find no indication for tunneling sidebands in $\delta$-AlOOH, in agreement with the $P$ evolution of the calculated potentials, showing no double-well character over the respective compression range.

\section{Equation-of-State}
\begin{figure*}[h!t]
\includegraphics[width=0.98\textwidth]{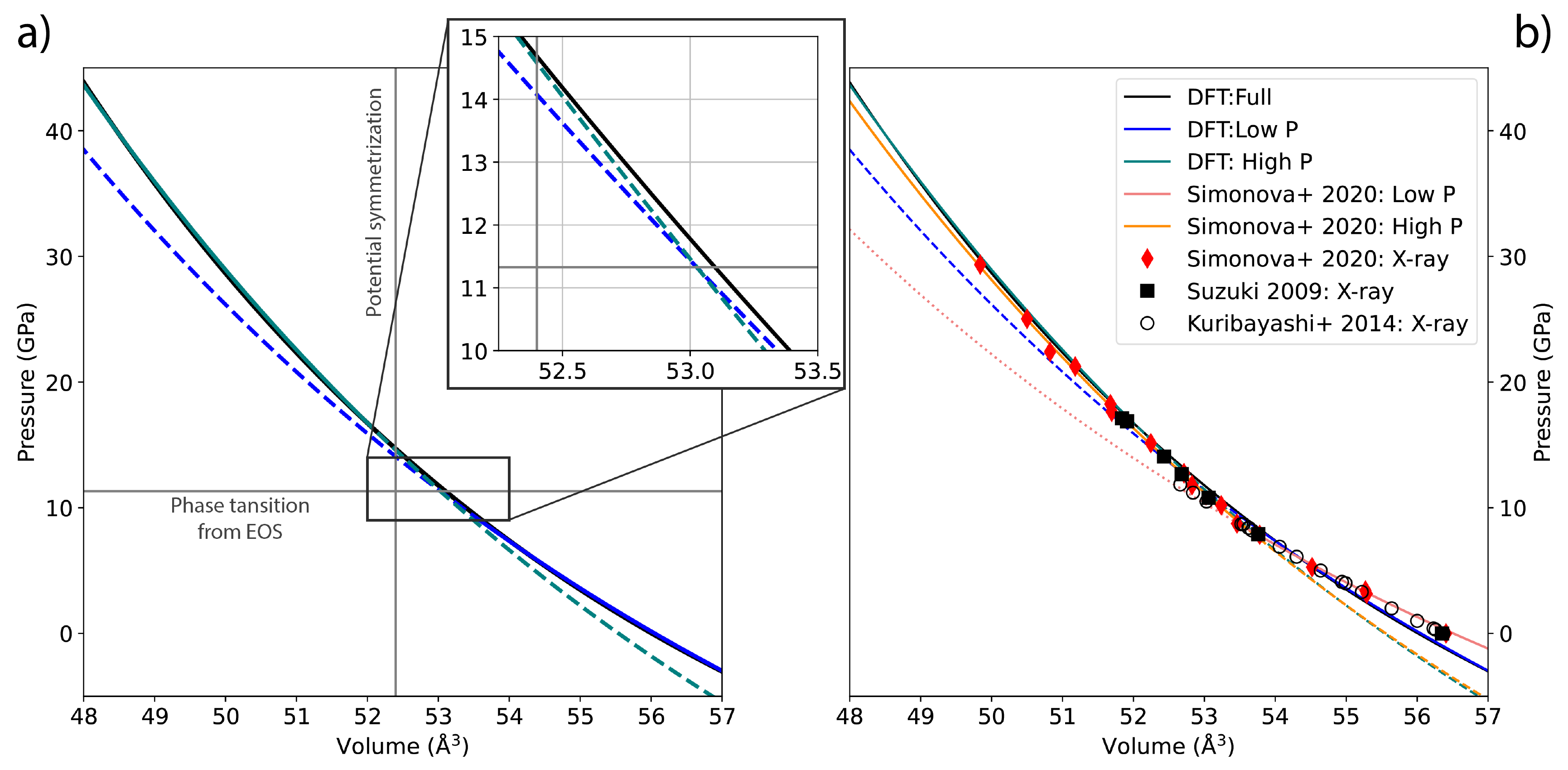}
\caption{\textbf{(a)} Third order Birch-Murnaghan (BM3) equation-of-state fit to the total energy of the optimized structures from the inverse configuration sampling. The full $V$-range is fitted by the black curve, $V> 53.5$ \AA$^3$ by the blue curve and $V<52$ \AA$^3$ by the teal curve. The vertical line shows the estimated transition from the potential analysis, and the horizontal line the intersection of the enthalpy calculated from the two partial BM3 fits. The inset shows a zoom of the area where low and high-$P$ partial EOS intersect. \textbf{(b)} Comparison of different $V-P$ data and equation-of-states from our calculations with literature data by \citet{simonova2020,kuribayashi2014} and \citet{suzuki2009compressibility}.}
\label{fig:transition}
\end{figure*}

In order to convert $V$ of the simulation cells to $P$, we use the optimized structures from the inverse configuration and fit a third order Birch-Murnaghan (BM3) EOS to total energy. We use the $V$ at which the potential symmetrizes (52.4 \AA$^3$) to split the $E-V$ results in two sets and calculate an EOS for (i) the full $V$-range, (ii) $57 \geq V \geq 53.5$ \AA$^3$ and (iii) $52 \geq V \geq 47$ \AA$^3$ (Fig.~\ref{fig:transition}a).
We calculate the intersection of the enthalpy curves for (ii) and (iii) and estimate the transition $P$ as $11.3 \pm  0.6 $ GPa (horizontal line in Fig.~\ref{fig:transition}a), where the error is estimated from the shift when the point closest to the transition in the potential is in-/excluded from the respective $V$ ranges. The phase transition is of second order as we do not find a $V$ collapse.

The calculated EOS parameters (low-$P$: $V_0= 56.0$  \AA$^3$, $K_0= 183$  GPa, $K_0^{'} =  3.7$ and high-$P$: $V_0 =  55.5$  \AA$^3$, $K_0 =  224$  GPa, $K_0^{'} =  4.0$) are in good agreement with partial EOS parameters by \citet{simonova2020} (low-$P$: $V_0=56.51(8)$ \AA$^3$, $K_0=142(5)$ GPa; high-$P$ part $V_0=55.56(8)$ \AA$^3$, $K_0=216.0(5)$ GPa with $K_{ \rm 0}^{'}=4.0$ in both cases)  and data from \citet{suzuki2009compressibility} and \citet{kuribayashi2014} at $V<55$  \AA$^3$ (Fig.~\ref{fig:transition}b). At larger $V$, our EOS slightly underestimates $P$  compared to the experimental EOS, which is most likely caused by thermal effects not being included in the calculations. 

Using the respective low-$P$ and high-$P$ EOS parameters, we convert all $V$ dependencies to $P$ dependencies and find potential symmetrization at $P=14.7\pm0.4$ GPa and the maxima in $\omega$ and $\omega^{'}$ at $P=13.1 \pm 0.7$ GPa (Table \ref{tab:sum_table}).

\begin{table*}
\begin{center}
\caption{Calculated values of the transition pressure from the low-$P$ $P2_1nm$ to the high-$P$ $Pnnm$ structure of AlOOH.  For the line width analysis,  errors are estimated from the $P$ error and for the chemical shift from the $P$ resolution of the measurement. For the simulations, errors are estimated from the $P$ resolution of the sampling and for the EOS calculation as the difference between the intersection $P$ of the low-$P$ and high-$P$ sub-EOS (Fig.~\ref{fig:transition}). The last three columns show literature data from \citet{simonova2020} (S+ 2020),  \citet{SF2018} (SF+ 2018) and \citet{pillai2018} (P+ 2018). }

\begin{tabular}{|c||c|c|c|c|c|c|c|c|}\hline
& & & & & & &&\\
Criterion~~&NMR Shift &~~~ EOS~~~&~~ $<$AlOAl$>$ angles ~~& NMR line width &~Potential~&~ S$+$ 2020~&~ SF$+$ 2018 ~&~P$+$ 2018 ~\\
& & & & & & &&\\\hline\hline
& & & & & & &&\\
$P$ (GPa) &$9.9 \pm 0.6 $& $11.3\pm 0.6$ & $13.1 \pm 0.7$ &$12.4 \pm 0.3$& $14.7 \pm 0.4 $&$> 10$ (structural)& 9/18.1&8/15 \\
& & & & & & &&\\ \hline
\end{tabular}
\label{tab:sum_table}
\end{center}
\end{table*}

\section{Discussion \& Conclusion}
We have analyzed different properties of $\delta$-AlOOH that can be directly linked to the phase transition (chemical shift, EOS) and hydrogen bond symmetrization (potential symmetry), but also properties that should record an influence of both processes (angles $\omega$ and $\omega^{'}$ as well as the NMR line width). We find features in the same three $P$ regions: (i) The phase transition characterized by the change in chemical shift at $9.9 \pm 0.6 $ GPa and the change in the slope of the $E-V$ curve (and therefore splitting of the EOS) at $11.3\pm 0.4$ GPa; (ii) a maximum in the angles $\omega$ and $\omega^{'}$ at $13.1 \pm 0.7$ GPa and a minimum in the line width at $12.4 \pm 0.3$ GPa; (iii) the symmetrization of the potential at  $14.7 \pm 0.4 $ GPa.

Comparing our results with recently published experiments \cite{SF2018,simonova2020} and computations \citep{cortona2017,pillai2018}, we find that we match the phase transition measured via neutron diffraction by \citet{SF2018}  and X-ray diffraction by \citet{simonova2020} at $\sim$ 10 GPa in the NMR experiments and the calculations. Furthermore, we reproduce the hydrogen bond symmetrization estimate from the calculation of elastic constants by \citet{cortona2017} and \citet{pillai2018} at $\sim15$ GPa in our potential analysis. For angles $\omega$ and $\omega^{'}$, we find the same increase in slope as \citet{SF2018}, with a maximum at $13.1 \pm 0.7$ GPa. If we take a closer look at the data by \citet{SF2018}, values for the angles $\omega$ and $\omega^{'}$ at the highest three $P$ points ($P \geq 13$ GPa)  appear to decrease linearly with $P$, in agreement with our computational results (Fig.~\ref{fig:Angles}). The large scatter and limited $P$-range in the experimental data impedes a more detailed analysis, comparison and discussion, however.


The absence of a double well in our mapping of the hydrogen potential at any $V$ is supported by low-field NMR measurements which do not show any indication of tunneling side bands. Therefore, we conclude that, contrary to the ice-VII to ice-X transition \cite{meier2018,trybel2020proton}, there is no tunneling-induced proton disorder in $\delta$-AlOOH. The only observation directly linked to a double-well potential are the Fourier difference maps by \citet{SF2018} that describe an asymmetric proton distribution at $P> 9.5$ GPa, followed by a bimodal distribution to $P \lesssim15$ GPa and a symmetric unimodal distribution at $P=18$ GPa. According to our results and data, the intermediate (bimodal) state is not characterized by proton disorder. Rather, it reflects order with weak asymmetry that gradually decreases. Therefore, we suggest the following interpretation of the neutron data: As O1 and O2 become symmetrically equivalent during the structural transition from $P2_1nm$ to $Pnnm$ at $\sim10$ GPa, asymmetry can no longer be associated to an O1-H$\cdots$O2 bond, but an averaged picture emerges where protons are closer to former O1 and O2 atoms, which may be visible in the Fourier difference maps and lead to the bimodal distribution if projected onto a $Pnnm$ unit cell.

\section*{\label{sec:Outlook} ACKNOWLEDGMENTS}
FT and GSN were supported by Deutsche Forschungsgemeinschaft (DFG) within FOR 2440 (Matter under Planetary Interior Conditions) with grant STE1105/13-1 and TM with grant ME5206/3-1. FT was further supported by the Swedish Research Council (VR) Grant No. 2019-05600.The authors thank Niccol\`o Satta and Giacomo Criniti for very helpful discussions. Computations were partly performed at the Leibniz Supercomputing Centre of the Bavarian Academy of Sciences and the Humanities. GPU accelerated computations are supported via the NVIDIA Hardware Grant. 
%

\end{document}